\documentclass[twoside,fleqn]{article}
\usepackage{amscd}
\usepackage{amsfonts}
\usepackage{epsf}
\usepackage{epic}
\usepackage{eepic}
\usepackage{latexsym}

\makeatletter
\def\fileversion{v2.6}
\def\filedate{24 November 1993}

\newdimen\@bls                    
\@bls=\baselineskip               
\advance\@bls -1ex                
\newdimen\@eps                    %
\def\section{\@startsection{section}{1}{\z@}
  {1.5\@bls plus 0.5\@bls}{1\@bls}{\normalsize\bf}}
\def\subsection{\@startsection{subsection}{2}{\z@}
  {1\@bls plus 0.25\@bls}{\@eps}{\normalsize\bf}}
\def\subsubsection{\@startsection{subsubsection}{3}{\z@}
  {1\@bls plus 0.25\@bls}{\@eps}{\normalsize\bf}}
\def\paragraph{\@startsection{paragraph}{4}{\parindent}
  {1\@bls plus 0.25\@bls}{0.5em}{\normalsize\bf}}
\def\subparagraph{\@startsection{subparagraph}{4}{\parindent}
  {1\@bls plus 0.25\@bls}{0.5em}{\normalsize\bf}}

\def\@sect#1#2#3#4#5#6[#7]#8{\ifnum #2>\c@secnumdepth
  \def\@svsec{}\else 
  \refstepcounter{#1}\edef\@svsec{\csname the#1\endcsname.\hskip0.5em}\fi
  \@tempskipa #5\relax
  \ifdim \@tempskipa>\z@
    \begingroup 
      #6\relax
      \@hangfrom{\hskip #3\relax\@svsec}{\interlinepenalty \@M #8\par}%
    \endgroup
    \csname #1mark\endcsname{#7}\addcontentsline
      {toc}{#1}{\ifnum #2>\c@secnumdepth \else
        \protect\numberline{\csname the#1\endcsname}\fi #7}%
  \else
    \def\@svsechd{#6\hskip #3\@svsec #8\csname #1mark\endcsname
      {#7}\addcontentsline{toc}{#1}{\ifnum #2>\c@secnumdepth \else
        \protect\numberline{\csname the#1\endcsname}\fi #7}}%
  \fi \@xsect{#5}}

\long\def\@makefigurecaption#1#2{\vskip 10mm #1. #2\par}

\long\def\@maketablecaption#1#2{\hbox to \hsize{\parbox[t]{\hsize}
  {#1 \\ #2}}\vskip 0.3ex}

\def\fnum@figure{Figure \thefigure}
\def\figure{\let\@makecaption\@makefigurecaption \@float{figure}}
\@namedef{figure*}{\let\@makecaption\@makefigurecaption \@dblfloat{figure}}

\def\table{\let\@makecaption\@maketablecaption \@float{table}}
\@namedef{table*}{\let\@makecaption\@maketablecaption \@dblfloat{table}}

\textfloatsep=\floatsep           
\intextsep=\floatsep              

\long\def\@makefntext#1{\parindent 1em\noindent\hbox{${}^{\@thefnmark}$}#1}

\def\maketitle{\begingroup        
    \def\thefootnote{\fnsymbol{footnote}}%
    \newpage \global\@topnum\z@ 
    \@maketitle \@thanks
  \endgroup
  \let\maketitle\relax \let\@maketitle\relax
  \gdef\@thanks{}\let\thanks\relax
  \gdef\@address{}\gdef\@author{}\gdef\@title{}\let\address\relax}

\def\justify@on{\let\\=\@normalcr
  \leftskip\z@ \@rightskip\z@ \rightskip\@rightskip}

\newbox\fm@box                    

\def\@maketitle{
  \global\setbox\fm@box=\vbox\bgroup
    \vskip 8mm                    
    \raggedright                  
    \hyphenpenalty\@M             
    {\Large \@title \par}         
    \vskip\@bls                   
    {\normalsize                  
     \@author \par}               
    \vskip\@bls                   
    \@address                     
  \egroup
  \twocolumn[
    \unvbox\fm@box                
    \vskip\@bls                   
    \unvbox\abstract@box          
    \vskip 2pc]}                  

\newcounter{address} 
\def\theaddress{\alph{address}}
\def\@makeadmark#1{\hbox{$^{\rm #1}$}}   

\def\address#1{\addressmark\begingroup
  \xdef\@tempa{\theaddress}\let\\=\relax
  \def\protect{\noexpand\protect\noexpand}\xdef\@address{\@address
  \protect\addresstext{\@tempa}{#1}}\endgroup}
\def\@address{}

\def\addressmark{\stepcounter{address}%
  \xdef\@tempb{\theaddress}\@makeadmark{\@tempb}}

\def\addresstext#1#2{\leavevmode \begingroup
  \raggedright \hyphenpenalty\@M \@makeadmark{#1}#2\par \endgroup
  \vskip\@bls}

\newbox\abstract@box              

\def\abstract{%
  \global\setbox\abstract@box=\vbox\bgroup
  \small\rm
  \ignorespaces}
\def\endabstract{\par \egroup}

\def\thebibliography#1{\section*{REFERENCES}\list{\arabic{enumi}.}
  {\settowidth\labelwidth{#1.}\leftmargin=1.67em
   \labelsep\leftmargin \advance\labelsep-\labelwidth
   \itemsep\z@ \parsep\z@
   \usecounter{enumi}}\def\makelabel##1{\rlap{##1}\hss}%
   \def\newblock{\hskip 0.11em plus 0.33em minus -0.07em}
   \sloppy \clubpenalty=4000 \widowpenalty=4000 \sfcode`\.=1000\relax}

\newcount\@tempcntc
\def\@citex[#1]#2{\if@filesw\immediate\write\@auxout{\string\citation{#2}}\fi
  \@tempcnta\z@\@tempcntb\m@ne\def\@citea{}\@cite{\@for\@citeb:=#2\do
    {\@ifundefined
       {b@\@citeb}{\@citeo\@tempcntb\m@ne\@citea
        \def\@citea{,\penalty\@m\ }{\bf ?}\@warning
       {Citation `\@citeb' on page \thepage \space undefined}}%
    {\setbox\z@\hbox{\global\@tempcntc0\csname b@\@citeb\endcsname\relax}%
     \ifnum\@tempcntc=\z@ \@citeo\@tempcntb\m@ne
       \@citea\def\@citea{,\penalty\@m}
       \hbox{\csname b@\@citeb\endcsname}%
     \else
      \advance\@tempcntb\@ne
      \ifnum\@tempcntb=\@tempcntc
      \else\advance\@tempcntb\m@ne\@citeo
      \@tempcnta\@tempcntc\@tempcntb\@tempcntc\fi\fi}}\@citeo}{#1}}

\def\@citeo{\ifnum\@tempcnta>\@tempcntb\else\@citea
  \def\@citea{,\penalty\@m}%
  \ifnum\@tempcnta=\@tempcntb\the\@tempcnta\else
   {\advance\@tempcnta\@ne\ifnum\@tempcnta=\@tempcntb \else
\def\@citea{--}\fi
    \advance\@tempcnta\m@ne\the\@tempcnta\@citea\the\@tempcntb}\fi\fi}

\def\ps@crcplain{\let\@mkboth\@gobbletwo
     \def\@oddhead{\reset@font{\sl\rightmark}\hfil \rm\thepage}%
     \def\@evenhead{\reset@font\rm \thepage\hfil\sl\leftmark}%
     \let\@oddfoot\@empty
     \let\@evenfoot\@oddfoot}

\ps@crcplain                    

\newcommand{\AmS}{{\protect\the\textfont2
  A\kern-.1667em\lower.5ex\hbox{M}\kern-.125emS}}

\makeatother

\newcommand{\bee}{\begin{equation}}
\newcommand{\ene}{\end{equation}}

\title{Heat-Kernel Asymptotics  of Locally Symmetric Spaces of Rank One
and Chern--Simons Invariants}

\author{{Andrei A. Bytsenko}
\address{Departamento de Fisica, Universidade Estadual de Londrina, 
Caixa Postal 6001, Londrina-Parana, Brazil}
\thanks{The author would like to thank Professor G. Esposito for giving
him opportunity to submit this invited contribution.}}

\begin{document}
\begin{abstract}
The asymptotic expansion of the heat kernel associated with Laplace 
operators is considered for general irreducible rank-one locally
symmetric spaces.
Invariants of the Chern--Simons theory of irreducible $U(n)-$ flat connections
on real compact hyperbolic 3-manifolds are derived. 
\end{abstract}

\maketitle

\section{Introduction}

The semiclassical approximation for the 
Chern--Simons partition function may be expressed by
the asymptotics which leads to a series of
$C^{\infty}-$ invariants associated with triplets $\{X;F;\xi\}$ with $X$ a 
smooth homology 
$3-$ sphere, $F$ a homology class of framings of $X$, and $\xi$ an acyclic 
conjugacy class of orthogonal representations of the fundamental group
$\pi_1(X)$ 
\cite{axel94-39-173}. In addition the cohomology $H(X;{\rm Ad}\,\xi)$ of $X$ 
with respect to the local system related to ${\rm Ad}\,\xi$ vanishes. 
In dimension three there are two important topological quantum field theories
of cohomological type, namely topological $SU(2)$ gauge theory of flat 
connection and a version of the Seiberg-Witten theory. The twisted 
${\mathcal N}=4$ $SUSY\,\,\, SU(2)$ pure gauge theory (version of the 
Donaldson-Witten
theory) describes the Casson invariant \cite{blau93} while Seiberg-Witten 
theory is a 3d twisted version of ${\mathcal N}=4$ $SUSY\,\,\, U(1)$ gauge 
theory with matter multiplet \cite{seib94,witt94}. Both theories can be 
derived from 4d ${\mathcal N}=2$ $SUSY\,\,\,SU(2)$ gauge theory corresponding 
via twist to Donaldson-Witten theory. It would be interesting and natural to 
investigate dual description of the ${\mathcal N}=2$ theory in low-energy
limit. It could provide formulation of invariants of four-manifolds involving
elements of the Chern--Simons invariants. In this paper we turn into 
Chern--Simons invariants related to locally symmetric spaces. 

The invariant $W_{CS}(X;k)$ associated with the Chern--Simons functional
$CS(A^{(j)})$ has to all orders in
$k^{-1}=\hbar/2\pi$ \,\,$(k\in {\Bbb Z})$ an asymptotic stationary phase 
approximation of the form \cite{roza96}

$$
\!\!\!\!\!\!\!\!\!\!\!\!\!\!\!\!\!\!\!\!\!\!\!\!
\!\!\!\!\!\!\!\!\!\!\!\!\!\!\!\!\!\!\!\!\!\!\!\!
\!\!\!\!
W_{CS}(X;k)=\sum_{j}W_{0}^{(j)}(X;k)
$$
$$
\times
\exp\,\sqrt{-1}k
\left(CS(A^{(j)})+\sum_{n=2}^{\infty}CS_n(A^{(j)})k^{-n}
\right)
\mbox{,}
\eqno{(1.1)}
$$
where $CS_n(A^{(j)})$ are the n-loop quantum corrections of flat connection
$A^{(j)}$ coming from the n-loop 1-particle irreducible Feynman diagrams.

The partition function of quadratic functional (one-loop expansion) 
$W_0(X;k)$ can be written in the form \cite{adam95u-95,adam98-417}

$$
W_{0}(X;k)
=\left(\frac{\hbar}{2}\right)^{\zeta(0,|{\frak D}|)/2}
e^{\left(-(\sqrt{-1}\pi/4)\eta(0,{\frak D})\right)}
$$
$$
\times
\left[T_{an}^{(2)}(X)\right]^{1/2}{\rm Vol}(X)^{-{\rm dim}H^{0}(\nabla)/2}
\mbox{.}
\eqno{(1.2)}
$$
The holomorphic function

$$
\!\!\!\!\!\!\!\!\!\!\!\!\!\!\!\!\!\!\!\!\!\!
\!\!\!\!\!\!\!\!\!\!\!
\eta(s,{\frak D})\stackrel{def}{=}
\sum_{\lambda\in {\rm Spec}\,{\frak D}\backslash\{0\}}
{\rm sgn}(\lambda)|\lambda|^{-s}
$$
$$
={\rm Tr}\left({\frak D}
\left({\frak D}^2\right)^{-(s+1)/2}\right)
\mbox{,}
\eqno{(1.3)}
$$
is well defined for all $\Re s\gg 0$ (in Eq. (1.3) the sum has to be taken over
all the spectrum $\lambda$) and extends to a meromorphic function
on ${\Bbb C}$.
Indeed, from the asymptotic behaviour of the heat kernel of the Dirac operator 
$\frak D$ at $t=0$,
${\rm Tr}\left({\frak D}\exp{(-t{\frak D}^2)}\right)={\mathcal O}(t^{1/2})$ 
\cite{bism86} 
it follows that $\eta(s,{\frak D})$ admits a meromorphic extension to the 
whole $s-$ plane, with at most simple poles at 
$s=({\rm dim}\,X-q)/({\rm ord}\,{\frak D})$\,\,\,$(q\in {\Bbb Z}_{+})$ and
locally computable residues.
It has been established that the point $s=0$ is not a pole, which makes it
possible to define the eta invariant of ${\frak D}$ by 
$\eta(0,{\frak D})$.
One can attach the eta invariant to any operator of Dirac type on a compact
Riemannian manifold of odd dimension. Dirac operators on even dimensional
manifolds have symmetric spectrums and, therefore, trivial eta invariants.
As far as the zeta function $\zeta(0,|{\frak D}|)$ is concerned in Eq. (1.2),
we recall that there exist $\varepsilon, \delta >0$ 
such that for $0<t<\delta$
the heat kernel expansion for self-adjoint Laplace operators ${\frak L}_p$
(acting on the space of $p-$ forms) is given by

$$
{\rm Tr}\left(e^{-t{\frak L}_p}\right)=\sum_{0\leq \ell \leq \ell_0}
A_{\ell}({\frak L}_p)t^{-\ell} +{\mathcal O}(t^{\varepsilon})
\mbox{.}
\eqno{(1.4)}
$$
We shall calculate the heat kernel coefficients $A_{\ell}$ for locally
symmetric spaces of rank one in the next sections.
One can shown that the zeta function $\zeta(s,|{\frak D}|)$ 
is well-defined and analytic for $\Re\,s>0$ and can be continued to a 
meromorphic function on ${\Bbb C}$, regular 
at $s=0$. Moreover (see Refs. \cite{adam95u-95,adam98-417}),

$$
\zeta(0,|{\frak D}|)=\sum_{p=0}(-1)^p(A_0({\frak L}_p)-{\rm dim} H^p(R(S)))
\mbox{,}
\eqno{(1.5)}
$$
where $R(S)$ is the resolvent of the quadratic functional $S$
(a chain of linear maps).
$\zeta(0,|{\frak D}|)$ can be expressed in terms of the dimensions of the 
cohomology spaces 
of ${\frak D}$. Indeed, for all 
$p$\, $A_0({\frak L}_p)=0$, because we are dealing with odd-dimensional 
manifold without boundary. Since 
$H^p(R(S_{\frak D}))=H^{m-p}(\nabla)$ (by virtue of Poincar{\`e} duality),
$m=({\rm dim}\,X-1)/2$, it follows that 

$$
\!\!\!\!\!\!\!\!\!\!\!\!\!\!\!\!\!\!\!\!\!\!\!\!
\zeta(0||{\frak D}|)=-\sum_{p=0}^m (-1)^p{\rm dim} H^p(R(S))
$$
$$
=(-1)^{m+1}
\sum_{p=0}^m (-1)^p{\rm dim} H^p({\nabla})
\mbox{.}
\eqno{(1.6)}
$$

The Ray-Singer norm $||\cdot||^{RS}$ on the determinant line
${\rm det}H(X;\xi)$ is defined by \cite{ray71-7}

$$
\!\!\!\!\!\!\!\!\!\!\!\!\!\!\!\!\!\!\!\!\!\!\!\!
\!\!\!\!\!\!\!\!\!\!\!\!\!\!\!\!\!\!\!\!\!\!\!\!
\!\!\!\!\!\!\!\!\!\!\!\!\!\!\!\!\!\!\!\!\!\!\!\!
\!\!\!\!\!\!\!\!\!\!\!\!\!\!\!\!\!\!\!\!\!\!\!\!
||\cdot||^{RS}\stackrel{def}=
$$
$$
|\cdot|\prod_{p=0}^{{\rm dim}X}
\left[\exp \left(-\frac{d}{ds}
\zeta (s,{\frak L}_p)|_{s=0}\right)\right]^{(-1)^pp/2}
\mbox{.}
\eqno{(1.7)}
$$
For a closed
connected orientable smooth manifold of odd dimension and for Euler structure
$\eta\in {\rm Eul}(X)$ the Ray-Singer norm of its cohomological torsion
$T_{an}^{(2)}(X;\eta)=T_{an}^{(2)}(X)\in {\rm det}H(X;\xi)$ is equal to the 
positive
square root of the absolute value of the monodromy of $\xi$ along the 
characteristic class $c(\eta)\in H^1(X)$ \cite{farb98}: 
$||T_{an}^{(2)}(X)||^{RS}=|{\rm det}_{\xi}c(\eta)|^{1/2}$. In the special 
case where the flat bundle $\xi$ is acyclic $(H^p(X;\xi)=0)$ we have

$$
\!\!\!\!\!\!\!\!\!\!\!\!\!\!\!\!\!\!\!\!\!\!\!\!\!\!\!\!\!\! 
\!\!\!\!\!\!\!\!\!\!\!\!\!\!\!\!\!\!\!\!\!\!\!\!\!\!\!\!\!\!
\left[T_{an}^{(2)}(X)\right]^2
=|{\rm det}_{\xi}c(\eta)|
$$
$$
\times
\prod_{p=0}^{{\rm dim}\,X}\left[\exp\left(-\frac{d}{ds}
\zeta (s,{\frak L}_p)|_{s=0}\right)\right]^{(-1)^{p+1}p}
\mbox{.}
\eqno{(1.8)}
$$

This note is an extension of previous  papers 
\cite{byts1,byts2,byts3,byts4,byts5,byts6,byts7,byts8}. Our aim is to evaluate
the semiclasssical partition function, weighted by $\exp[\sqrt{-1}kCS(A)]$. 
We shall do this analysis using the spectral properties of elliptic operators
acting on locally symmetric spaces of rank one.

\section{Asymptotics of the heat kernel on rank one locally symmetric spaces}

In Refs.  \cite{miat76,miat79-29-249,miat80-260-1}, Miatello studies the 
case of a closed locally
symmetric rank one manifold $X$, using the representation theory of the group
of isometries of $X$. We consider the same case, but
we use the spectral zeta function of $X$. By our approach we determine the
expansion coefficients explicitly,
given the results of Ref. \cite{will98-182-137}.
We shall be working with an irreducible rank one symmetric space $M=G/K$
of non-compact type. Thus $G$ will be a connected non-compact simple split rank
one Lie group with finite centre and $K\subset G$ will be a maximal compact
subgroup \cite{helg62}. Let $\Gamma\subset G$ be a discrete,
co-compact torsion free subgroup. Then $X=X_{\Gamma}=\Gamma\backslash M$ is a 
compact Riemannian manifold with fundamental group $\Gamma$, i.e. $X$ is a 
compact locally symmetric space. Given a finite-dimensional unitary 
representation $\chi$ of $\Gamma$ there is the corresponding vector bundle 
$V_{\chi}\rightarrow X$ over $X$ given by $V_{\chi}=\Gamma\backslash (M\otimes 
F_{\chi})$, where $F_{\chi}$ (the fibre of $V_{\chi}$) is the representation 
space of $\chi$ and where $\Gamma$ acts on $M\otimes F_{\chi}$ by the rule
$\gamma\cdot(m,f)=(\gamma\cdot m,\chi(\gamma)f)$ for $(\gamma,m,f)\in (\Gamma
\otimes M\otimes F_{\chi})$. Let ${\frak L}_{\Gamma}$ be the Laplace-Beltrami
operator of $X$ acting on smooth sections of $V_{\chi}$; we obtain 
${\frak L}_{\Gamma}$ by projecting the Laplace-Beltrami operator of $M$ 
(which is
$G-$ invariant and thus $\Gamma-$ invariant) to $X$. 

The spectral zeta function 
$\zeta(s;{\frak L}_{\Gamma})\equiv\zeta_{\Gamma}(s;\chi)$ of $X_{\Gamma}$ of
Minakshisundaram-Pleijel type \cite{mina49-1-242} 
is a holomorphic function on the
domain $\Re s>d/2$, where $d$ is the dimension of $M$, and by general 
principles $\zeta_{\Gamma}(s;\chi)$ admits a meromorphic continuation to the
full complex plane ${\Bbb C}$. However since the manifold $X_{\Gamma}$ is
quite special it is desirable to have the meromorphic continuation of
$\zeta_{\Gamma}(s;\chi)$ in an explicit form, for example in terms of the
structure of $G$ and $\Gamma$. Using the Selberg trace formula and the $K$-
spherical harmonic analysis of $G$, such a form has been obtained in 
\cite{will98-182-137}; also see Refs. \cite{rand75-201-241,will97-38-796}. 
To state these results we introduce further notation.

Up to local isomorphism we can represent $M=G/K$ by the following quotients:

$$
M=\left[ \begin{array}{ll}SO_1(n,1)/SO(n)\,\,\,\,\,\,\,\,\,\,\,\,\,\,\,\,\,
\,\,\,\,\,\,\,\,\,\,\,\,\,\,\,\,\,\,\,\,\,(I) \\
SU(n,1)/U(n)\,\,\,\,\,\,\,\,\,\,\,\,\,\,\,\,\,\,\,\,\,\,\,\,\,\,\,\,\,\,\,\,
\,\,\,\,\,\,\,\,\,(II)
\\SP(n,1)/(SP(n)\otimes SP(1))\,\,\,\,\,(III)\\
F_{4(-20)}/Spin(9)\,\,\,\,\,\,\,\,\,\,\,\,\,\,\,\,\,\,\,\,\,\,\,\,\,\,\,\,\,
\,\,\,\,\,\,(IV)
\end{array} \right]
\mbox{}
\eqno{(2.1)}
$$
where $d=n, 2n, 4n, 16$ and $\rho_0$ is given by 
$\rho_0=(n-1)/2, n, 2n+1, 11$ respectively in the cases
$(I)$ to $(IV)$. For details on these matters the reader may consult 
\cite{helg62}, and also the Appendix in \cite{will97-38-796}.
The spherical harmonic analysis on $M$ is controlled by Harish-Chandra's 
Plancherel density $\mu (r)$, a function on the real numbers $\Bbb R$,
computed by Miatello \cite{miat76,miat79-29-249,miat80-260-1}, and others, in 
the rank one case we are
considering.

\subsection{The heat kernel coefficients} 
The object of interest is the heat 
kernel $\omega_{\Gamma}(t;\chi)$ defined for $t>0$ by

$$
\omega_{\Gamma}(t;\chi)=\sum_{j=0}^{\infty}n_j(\chi)
e^{-\lambda_j(\chi)t}
\mbox{,}
\eqno{(2.2)}
$$
where $n_j$ is the (finite) multiplicity of the eigenvalue $\lambda_j$.
If $h_t$ is the fundamental solution of the heat equation on $M$, then
$h_t$ and $\omega_{\Gamma}(t;\chi)$ are related by the Selberg trace
formula (cf. \cite{will98-182-137})

$$
\omega_{\Gamma}(t;\chi)=\chi(1)\mbox{Vol}(\Gamma\backslash G)h_t(1)
+\theta_{\Gamma}(t;\chi)
\mbox{,}
\eqno{(2.3)}
$$
where we denote by ${\rm Vol}(\Gamma\backslash G)$ the $G-$ invariant volume of 
$\Gamma\backslash G$ induced by Haar measure on $G$,
the theta function $\theta_{\Gamma}(t;\chi)$ is given by Eq. (4.18) of
\cite{will98-182-137} (for $b=0$ there) and where

$$
h_t(1)=\frac{1}{4\pi}e^{-\rho_0^2t}\int_{{\Bbb R}}
e^{-r^2t}\mu (r)dr
\mbox{.}
\eqno{(2.4)}
$$
We shall {\em not} need the result (2.2). Our goal is to compute explicitly
all coefficients $A_k=A_k(\Gamma,\chi)$ in the asymptotic 
expansion

$$
\omega_{\Gamma}(t;\chi)\simeq (4\pi t)^{-d/2}\sum_{k=0}^{\infty}A_kt^k,
\,\,\,\,\,\mbox {as} \,\, t\rightarrow 0^{+}
\mbox{.}
\eqno{(2.5)}
$$
$\zeta_{\Gamma}(s;\chi)$ and $\omega_{\Gamma}(t;\chi)$ are related by the
Mellin transform:

$$
\!\!\!\!\!\!\!\!\!\!\!\!\!\!\!\!\!\!\!\!\!\!\!\!\!\!\!\!\!\!\!\!
\!\!\!\!\!\!\!\!\!\!\!\!\!\!\!\!\!\!\!\!\!\!\!\!\!\!\!\!\!\!\!\!
\!\!\!\!\!
\zeta_{\Gamma}(s;\chi)=\frac{{\frak M}[\omega_{\Gamma}](s)}{\Gamma(s)}
$$
$$
=
\frac{1}{\Gamma(s)}\int_0^{\infty}\omega_{\Gamma}(t;\chi)t^{s-1}dt,\,\,\,\,\,
\mbox{for}\,\,\,\, \Re s>\frac{d}{2}
\mbox{.}
\eqno{(2.6)}
$$
Moreover one knows by abstract generalities 
(cf. \cite{mina49-1-242,voro87-110-439} for example)
that the coefficients $A_k$ are related to residues and special values of
$\zeta_{\Gamma}(s;\chi)$. We obtain the following main result.

{\bf Theorem 2.1} (Ref. \cite{byts5}). {\em  
The heat kernel $\omega_{\Gamma}(t;\chi)$ in (2.2) admits an asymptotic
expansion (2.5). For all $G$ except $G=SO_1(\ell,1)$, $SU(q,1)$ with 
$\ell$ odd and $q$ even, and for $0\leq k\leq d/2-1$,

$$
A_k(\Gamma,\chi) = (4\pi)^{\frac{d}{2}-1}\chi(1){\rm Vol}
(\Gamma\backslash G)C_G\pi
$$
$$
\sum_{\ell=0}^k
\frac{(-\rho_0^2)^{k-\ell}}{(k-\ell)!}\Bigr[\frac{d}{2}-(\ell+1)\Bigr]!
a_{2[\frac{d}{2}-(\ell+1)]}
\mbox{,}
\eqno{(2.7)}
$$ 
while for $n=0,1,2,...$ we have

$$
A_{\frac{d}{2}+n}(\Gamma,\chi) = (-1)^n(4\pi)^{\frac{d}{2}-1}\chi(1)
{\rm Vol}(\Gamma\backslash G)C_G\pi
$$
$$
\times
\left[\sum_{j=0}^{\frac{d}{2}-1}(-1)^{j+1}\frac{\rho_0^{2(n+1+j)}j!a_{2j}}
{(n+1+j)!} \right. 
$$
$$
\left.
+ 2\sum_{j=0}^{\frac{d}{2}-1}\sum_{\ell=0}^n(-1)^{\ell}
\frac{\rho_0^{2(n-\ell)}}{(n-\ell)!}\beta_{\ell+1}(j)a_{2j}\right] 
\mbox{.}
\eqno{(2.8)} 
$$
Here $\beta_r(j)\,\,(r\in {\Bbb Z}_{+})$ is given by 

$$
\beta_r(j)\stackrel{def}{=}\left[2^{1-2(r+j)}-1\right]\left[\frac{\pi}{a(G)}
\right]^{2(r+j)}
$$
$$
\times
\frac{(-1)^jB_{2(r+j)}}{2(r+j)[(r-1)!]}
\mbox{,}
\eqno{(2.9)}
$$
$B_r$ is the $r$-th Bernoulli number, 

$$
a(G)\stackrel{def}{=}\left[ \begin{array}{ll}\pi \hspace{0.3cm}
\mbox{if\,\, $G=SO_1(\ell,1)$ with $\ell$ even,}\\
\frac{\pi}{2} \hspace{0.3cm}\mbox{if\,\, $G=SU(q,1)$\,\,\, with $q$ odd,}\\
\hspace{0,5cm}\mbox{or $G=SP(\ell,1)$ any $\ell$, $F_{4(-20)}$}\\
\end{array} \right]
\eqno{(2.10)}
$$
and $a_{2j},\, C_G$ are some constants ($C_G$ depending on $G$). 
For $G=SO_1(2n+1,1),\,\, k=0,1,2, ...$
   
$$
A_k(\Gamma,\chi)=\pi(4\pi)^{n-\frac{1}{2}}\chi(1){\rm Vol}
(\Gamma\backslash G)C_G
$$
$$
\times
\sum_{\ell=0}^{{\rm min}(k,n)}
\frac{(-n^2)^{k-\ell}\Gamma\left(n-\ell+\frac{1}{2}\right)a_{2(n-\ell)}}
{(k-\ell)!}
\mbox{.}
\eqno{(2.11)}
$$
}

\subsection{The case $G=SU(q,1)$} 

We considered the Minakshisundaram-Pleijel coefficients
$A_k(X_{\Gamma})$ for all compact rank one space forms $X_{\Gamma}$ 
(up to local isomorphism) with one exception - i.e. the case 
$X_{\Gamma}= \Gamma\backslash G/K$ with $G=SU(q,1)$. 
We assume $G=SU(q,1)$ where now $q\geq 2$ is even. The meromorphic structure of
$\zeta_{\Gamma}(s;\chi)$ in this case differs essentially from the 
case of odd $q$ in its non-singular terms - not the singular terms of
$\zeta_{\Gamma}(s;\chi)$ where information on poles is determined \cite{byts5}. 
We now define $\beta_r(j)$ by

$$
\beta_r(j)=\frac{(-1)^j2^{2(r+j)}B_{2(r+j)}}
{2(r+j)[(r-1)!]^{-1}}
\mbox{.}
\eqno{(2.12)}
$$
At this point the earlier discussions
apply and we may conclude the following.

{\bf Theorem 2.2}. {\em Formulae (2.7) and (2.8) also hold 
for $G=SU(q,1)$ with $q\geq 2$ even,
where $d/2=q=\rho_0$, provided that in formula (2.8) definition (2.9)
for $\beta_r(j)$ is replaced by definition (2.12).}

\section{Real compact hyperbolic manifolds}

In this section we consider briefly the Freed trace formula which is useful 
for calculation of the partition function of quadratic functional defined
on real hyperbolic space. The heat kernel coefficients for $p$-forms on 
hyperbolic space can be found in the paper of 
F.L. Williams in this volume.    

As before $X_{\Gamma}=\Gamma\backslash G/K$ is a 
compact manifold,
$G=SO_1(n,1)$ \,$(n\in {\Bbb Z}_{+})$ and  $K=SO(n)$. 
The corresponding symmetric space of non-compact type is the real hyperbolic
space ${\Bbb H}^n$ of sectional curvature $-1$. Its compact dual space is
the unit $n-$ sphere.

\subsection{Fried's trace formula} 

Let $a_0, n_0$
denote the Lie algebras of $A, N$ in an Iwasawa decomposition $G=KAN$. 
Since the rank of $G$ is one,
$\dim a_0=1$ by definition, say $a_0={\Bbb R}H_0$ for a suitable basis vector
$H_0$. One can normalize the choice of $H_0$ by $\beta(H_0)=1$, where
$\beta: a_0\rightarrow{\Bbb R}$ is the positive root which defines $n_0$; 
for more
detail see Ref. \cite{will97-38-796}. Since $\Gamma$ is torsion free, each
$\gamma\in\Gamma-\{1\}$ can be represented uniquely as some power of a 
primitive
element $\delta:\gamma=\delta^{j(\gamma)}$ where $j(\gamma)\geq 1$ is an 
integer and
$\delta$ cannot be written as $\gamma_1^j$ for $\gamma_1\in \Gamma$, \,\,
$j>1$ an
integer. Taking $\gamma\in\Gamma$, $\gamma\neq 1$, one can find $t_\gamma>0$
and $m_{\gamma}\in {\frak M} \stackrel{def}{=}\{m_{\gamma}\in K | m_{\gamma}a=
am_{\gamma}, \forall a\in A\}$ such that $\gamma$
is $G$ conjugate to $m_\gamma\exp(t_\gamma H_0)$, namely for some
${\rm g}\in G, \,{\rm g}\gamma {\rm g}^{-1}=m_\gamma\exp(t_\gamma H_0)$. 
Besides let 
$\chi_{\sigma}(m) = {\rm trace}(\sigma(m))$ be the character of $\sigma$,
for $\sigma$ a finite-dimensional representation of ${\frak M}$.

For $0\leq p\leq d-1$ the Fried trace formula holds \cite{fried}:

$$
{\rm Tr}\left(e^{-t{\cal L}^{(p)}}\right)=I^{(p)}(t,b^{(p)})
+I^{(p-1)}(t,b^{(p-1)})
$$
$$
+H^{(p)}(t,b^{(p)})+
H^{(p-1)}(t,b^{(p-1)})
\mbox{,}
\eqno{(3.1)}
$$
where
$$
I^{(p)}(t,b^{(p)})\stackrel{def}{=}\frac{\chi(1){\rm Vol}
(\Gamma\backslash G)}{4\pi}
$$
$$
\times
\int_{\Bbb R}\mu_{\sigma_p}(r)e^{-t[r^2+b^{(p)}+(\rho_0-p)^2]}dr
\mbox{,}
\eqno{(3.2)}
$$
$$
H^{(p)}(t,b^{(p)})\stackrel{def}{=}\frac{1}{\sqrt{4\pi t}}
\sum_{\gamma\in C_
\Gamma-\{1\}}\frac{\chi(\gamma)}{j(\gamma)}t_\gamma C(\gamma)
$$
$$
\times
\chi_{\sigma_p}
(m_\gamma)
\exp\left\{-\left[b^{(p)}t+(\rho_0-p)^2t+
\frac{t_\gamma^2}{4t}\right]\right\}
\mbox{,}
\eqno{(3.3)}
$$
$\rho_0=(d-1)/2$, $b^{(p)}$ are some constants, and the function 
$C(\gamma)$,\,\, $\gamma\in\Gamma$, defined on $\Gamma-\{1\}$ by

$$
C(\gamma)\stackrel{def}=e^{-\rho_0t_\gamma}|\mbox{det}_{n_0}\left(\mbox{Ad}
(m_\gamma
e^{t_\gamma H_0})^{-1}-1\right)|^{-1}\mbox{.}
\eqno{(3.4)}
$$
For $\mbox{Ad}$ denoting the
adjoint representation of $G$ on its complexified Lie algebra, one can compute
$t_\gamma$ as follows \cite{wall76-82-171}:

$$
e^{t_\gamma}=\mbox{max}\{|c||c= \mbox{an eigenvalue of}\,\, \mbox{Ad}(\gamma)\}
\mbox{.}
\eqno{(3.5)}
$$
Here $C_{\Gamma}$ is a complete set of representatives in $\Gamma$ of its
conjugacy classes; Haar measure on $G$ is suitably normalized.
For $p=0$ (i.e. for smooth
functions or smooth vector bundle sections) the measure 
$\mu(r)\equiv \mu_{0}(r)$ corresponds to the trivial representation of ${\frak M}$. 
For $p\geq 1$ there is a measure $\mu_{\sigma}(r)$ corresponding to a 
general irreducible representation $\sigma$ of ${\frak M}$. 
Let $\sigma_p$ be the 
standard representation of ${\frak M}=SO(d-1)$ on $\Lambda^p{\Bbb C}^{(d-1)}$. If
$d=2n$ is even then $\sigma_p\,\,(0\leq p\leq d-1)$ is always irreducible; if
$d=2n+1$ then every $\sigma_p$ is irreducible except for $p=(d-1)/2=n$, in 
which case $\sigma_n$ is the direct sum of two spin-$(1/2)$ representations 
$\sigma^{\pm}:\,\,\sigma_n=\sigma^{+}\oplus\sigma^{-}$. 
For $p=n$ the representation $\tau_n$ of $K=SO(2n)$ on 
$\Lambda^n {\Bbb C}^{2n}$ is not irreducible, 
$\tau_n=\tau_n^{+}\oplus\tau_n^{-}$ is the direct sum of spin-$(1/2)$ 
representations.

\subsection{The Harish-Chandra Plancherel measure} 

We should note that the reason for the pair of terms 
$\{I^{(p)}, I^{(p-1)}\}$,
\,\, $\{H^{(p)}, H^{(p-1)}\}$ in the trace formula 
Eq. (3.1) is that $\tau_p$ satisfies $\tau_p|_{{\frak M}}=
\sigma_p\oplus\sigma_{p-1}$. Then we have

$$
\mu_{\sigma_p}(r)= C^{(p)}(d)P(r,d)\times\left\{ \begin{array}{ll}
\tanh (\pi r), & \mbox{$d=2n$}\\
1, & \!\!\!\!\!\!\!\!
\mbox{$d=2n+1$}
\end{array}
\right.
$$
$$
=C^{(p)}(d)\times\left\{ \begin{array}{lll}

\sum\!_{\ell=0}^{d/2-1}\,\,
a_{2\ell}^{(p)}(d)r^{2\ell+1}\tanh (\pi r) & 
\,\,\, \mbox{}\\
\,\,\,\,\,\\
\sum\!_{\ell=0}^{(d-1)/2}\,\,a_{2\ell}^{(p)}(d)r^{2\ell} & \,\,\,
\mbox{}
\end{array}
\right.
\!\!\!\!\!\!\!\!
\mbox{,}
\eqno{(3.6)}
$$
$$
C^{(p)}(d)=\left(\begin{array}{c}
d-1 \\
p
\end{array}\right)\frac{\pi}{2^{2d-4}\Gamma(d/2)^2}\,\,
\mbox{,}
\eqno{(3.7)}
$$
where the $P(r,d)$ are
even polynomials (with suitable coefficients $a_{2\ell}^{(p)}(d)$) of degree 
$d-1$ for
$G\neq SO_1(2n+1,1)$, and of degree $d=2n+1$ for $G=SO_1(2n+1,1)$
\cite{eliz94,byts96,will97-38-796}.

\subsection{Case of the trivial representation}
For $p=0$ we take $I^{(-1)}
=H^{(-1)}=0$. Since $\sigma_0$ is the trivial representation one has
$\chi_{\sigma_0}(m_{\gamma})=1$. In this case formula (2.3)
reduces exactly to the trace formula for $p=0$ \cite{wall76-82-171,will90-242}:

$$
\!\!\!\!\!\!\!\!\!\!\!\!\!\!\!\!\!\!\!\!\!\!\!\!\!\!\!\!\!\!
\!\!\!\!\!\!\!\!\!\!\!\!\!\!
\omega_{\Gamma}^{(0)}(t,b^{(0)})=\frac{\chi(1)\mbox{vol}(\Gamma\backslash G)}
{4 \pi}
$$
$$
\times
\int_{\Bbb R}\mu_{\sigma_0}(r)e^{-(r^2+b^{(0)}+\rho_0^2)t}dr+
H^{(0)}(t,b^{(0)})
\mbox{,}
\eqno{(3.8)}
$$
where $\rho_0$ is associated with the positive restricted
(real) roots of $G$ (with multiplicity) with respect to a nilpotent factor $N$
of $G$ in an Iwasawa decomposition $G=KAN$. The function $H^{(0)}(t,b^{(0)})$ 
has the form

$$
H^{(0)}(t,b^{(0)})=\frac{1}{\sqrt{4\pi t}}
\sum_{\gamma\in C_
\Gamma-\{1\}}\chi(\gamma)t_\gamma j(\gamma)^{-1}
$$
$$
\:\:\:\:\:\:\:\:\:\:\:\:\:\:\:\:\:\:\:\:\:\:\:\:\:\:\:\:\:\:\:\:
\times
C(\gamma)e^{-[b^{(0)}t+
\rho_0^2t+t_\gamma^2/(4t)]}
\mbox{.}
\eqno{(3.9)}
$$

\subsection{Case of zero modes}
It can be shown \cite{will98-182-137} that the Mellin transform of 
$H^{(0)}(t,0)$ ($b^{(0)}=0$, i.e. the zero modes case)

$$
{\frak H}^{(0)}(s)\stackrel{def}{=}\int_0^{\infty}H^{(0)}(t,0)
t^{s-1}dt\mbox{,}
\eqno{(3.10)}
$$
is a holomorphic function on the domain $\Re s<0$. Then using the result of
Refs. \cite{byts96,will97-38-796} one can obtain on $\Re s<0$,

$$
\!\!\!\!\!\!\!\!\!\!\!\!\!\!\!\!\!\!\!\!\!\!\!\!\!\!\!\!\!\!\!\!\!\!\!
\!\!\!\!\!\!\!\!\!\!\!\!\!\!\!\!\!\!\!\!\!\!\!\!\!\!\!\!\!\!
{\frak H}^{(0)}(s)=\frac{\sin (\pi s)}{\pi}\Gamma(s)
$$
$$
\times
\int_0^{\infty}
\psi_\Gamma(t+2\rho_0;\chi)(2\rho_0t+t^2)^{-s}dt
\mbox{.}
\eqno{(3.11)}
$$
Here $\psi_\Gamma(s;\chi)\equiv d(\mbox{log}Z_\Gamma(s;\chi))/ds$,\,\,\, and
$Z_\Gamma(s;\chi)$ is a meromorphic suitably normalized Selberg zeta function
\cite{selb,fr,gang,will90-242,will,byts96}.

\section{The index theorem and the contribution to the partition 
function}

For any representation $\chi: \Gamma\rightarrow U(n)$ one can construct a 
vector bundle ${\Bbb {\widetilde   E}}_{\chi}$ over a certain 4-manifold 
$Y$ with 
boundary $\partial Y=X$ which is an extension of a 
flat vector bundle ${\Bbb E}_{\chi}$ over $X$.
Let ${\widetilde   A}_{\chi}$ be any extension of a flat connection 
$A_{\chi}$
corresponding to $\chi$. The index theorem of Atiyah-Patodi-Singer for the 
twisted Dirac operator $D_{{\widetilde   A}_{\chi}}$ 
\cite{atiy75-77,atiy75-78,atiy76-79} is given by

$$
\!\!\!\!\!\!\!\!\!\!\!\!\!\!\!\!\!\!\!\!\!\!\!\!
\!\!\!\!\!\!\!\!\!\!\!\!\!\!\!
{\rm Index}\left(D_{{\widetilde A}_{\chi}}\right)=\int_{Y}{\rm ch}
({\Bbb {\widetilde E}}_{\chi})
{\widehat A}(Y)
$$
$$
-\frac{1}{2}(\eta(0,{\frak D}_{\chi})+
h(0,{\frak D}_{\chi}))
\mbox{,}
\eqno{(4.1)}
$$
where ${\rm ch}({\Bbb {\widetilde   E}}_{\chi})$ and ${\widehat  A}(Y)$
are the Chern character and ${\widehat  A}-$ genus respectively,
${\widehat  A}=1-p_1(Y)/24,\,p_1(Y)$ is the 1-st Pontryagin class,
$h(0,{\frak D}_{\chi})$ is the dimension of the space of harmonic 
spinors on $X_{\Gamma}$ ($h(0,{\frak D}_{\chi})
={\rm dim}{\rm ker}\,{\frak D}_{\chi}$ = 
multiplicity of the 0-eigenvalue of ${\frak D}_{\chi}$ acting on 
$X$); ${\frak D}_{\chi}$ is a Dirac operator on 
$X$ acting on spinors with coefficients in $\chi$. 

The Chern--Simons invariants of $X$ can be derived
from Eq. (4.1). 
Indeed we have (see for detail Refs. \cite{byts3,byts4,byts7,byts8})):

$$
CS(\chi)\equiv 
\frac{1}{2}\left({\rm dim}{\chi}\eta(0,{\frak D})-
\eta(0,{\frak D}_{\chi})\right)
\,\,\,\,{\rm mod}({\Bbb Z}/2)
\mbox{.}
\eqno{(4.2)}
$$

A remarkable formula relating $\eta(s,{\frak D})$, to the closed geodesics 
on $X=\Gamma\backslash {\Bbb H}^3$ 
has been derived in \cite{mill78-108,mosc89}. More explicitly the following 
function can be defined, initially for $\Re(s^2)\gg 0$, by the formula

$$
\!\!\!\!\!\!\!\!\!\!\!\!\!\!\!\!\!\!\!\!\!\!\!\!\!\!\!\!\!\!\!\!
\!\!\!\!\!\!\!\!\!\!\!\!\!\!\!\!\!\!\!\!\!\!\!\!\!\!\!\!\!\!\!\!
\!\!\!\!\!\!\!\!\!\!\!\!\!\!\!\!\!\!\!
{\rm log}{\widetilde Z}(s,{\frak D})\stackrel{def}{=}
$$
$$
\sum_{[\gamma]\in {\mathcal  E}_1(\Gamma)}
(-1)^q\frac{L(\gamma,{\frak D})}
{|{\rm det}(I-P_h(\gamma))|^{1/2}}\frac{e^{-s\ell(\gamma)}}{m(\gamma)}
\mbox{,}
\eqno{(4.3)}
$$
where ${\mathcal  E}_1(\Gamma)$ is the set of those conjugacy classes 
$[\gamma]$
for which $X_{\gamma}$ has the property that the Euclidean de Rham factor
of ${\widetilde X}_{\gamma}$ is 1-dimensional (${\widetilde X}$ is a simply
connected cover of $X$ which is a symmetric space of noncompact type), the 
number $q$ is half the
dimension of the fibre of the centre bundle $C(TX)$ over $X_{\gamma}$, and 
$L(\gamma,{\frak D})$ is the Lefschetz number (see Ref. \cite{mosc89}).
Furthermore ${\rm log}{\widetilde Z}(s,{\frak D})$ has a meromorphic 
continuation to ${\Bbb C}$ given by the identity

$$
{\rm log}{\widetilde Z}(s,{\frak D})={\rm log}{\rm det}
\!\!
\left(\frac{{\frak D}-\sqrt{-1}s}{{\frak D}+\sqrt{-1}s}\right)
$$
$$
\:\:\:\:\:\:\:\:\:
+\sqrt{-1}\pi\eta(s,{\frak D})
\mbox{,}
\eqno{(4.4)}
$$
where $s\in \sqrt{-1}({\rm Spec}({\frak D})-\{0\})$, and 
${\widetilde Z}(s,{\frak D})$ 
satisfies the functional equation

$$
{\widetilde Z}(s,{\frak D}){\widetilde Z}(-s,{\frak D})=
e^{2\pi \sqrt{-1}\eta(s,{\frak D})}
\mbox{.}
\eqno{(4.5)}
$$
Let now $\chi: \Gamma\rightarrow U(F)$ be a unitary representation of 
$\Gamma$ on $F$. The Hermitian vector bundle 
${\Bbb F}={\widetilde X}\times_{\Gamma}F$ over $X$ inherits a flat 
connection from the trivial connection on
${\widetilde X}\times F$. We specialize to the case of locally 
homogeneous Dirac operators 
${\frak D}: C^{\infty}(X,{\Bbb E})\rightarrow C^{\infty}(X,{\Bbb E})$ in
order to construct a generalized operator ${\mathcal  O}_{\chi}$, acting on
spinors with coefficients in $\chi$. 
If ${\frak D}: C^{\infty}(X,V)\rightarrow C^{\infty}(X,V)$ is
a differential operator acting on the sections of the vector bundle $V$,
then ${\frak D}$ extends canonically to a differential operator 
${\frak D}_{\chi}: C^{\infty}(X,V\otimes{\Bbb F})\rightarrow 
C^{\infty}(X,V\otimes{\Bbb F})$, uniquely characterized by the property that 
${\frak D}_{\chi}$ is locally isomorphic to 
${\frak D}\otimes...\otimes {\frak D}$\,\,\,(${\rm dim}\,F$ times) 
\cite{mosc89}.
 
One can repeat the arguments to construct a twisted
zeta function ${\widetilde Z}(s,{\frak D}_{\chi})$. There exists a
zeta function  ${\widetilde Z}(s,{\frak D}_{\chi})$, meromorphic on ${\Bbb C}$, 
given for $\Re(s^2)\gg 0$ by the formula

$$
\!\!\!\!\!\!\!\!\!\!\!\!\!\!\!\!\!\!\!\!\!\!\!\!\!\!\!\!\!\!\!\!
\!\!\!\!\!\!\!\!\!\!\!\!\!\!\!\!\!\!\!\!\!\!\!\!\!\!\!\!\!\!\!\!
\!\!\!\!\!\!\!\!\!\!\!\!\!\!\!\!\!\!\!
{\rm log}{\widetilde Z}(s,{\frak D}_{\chi})\stackrel{def}{=}
$$
$$
\sum_{[\gamma]\in {\mathcal  E}_1(\Gamma)}
(-1)^q{\rm Tr}\chi(\gamma)\frac{L(\gamma,{\frak D})}
{|{\rm det}(I-P_h(\gamma))|^{1/2}}\frac{e^{-s\ell(\gamma)}}{m(\gamma)}
\mbox{;}
\eqno{(4.6)}
$$
moreover one has ${\rm log}{\widetilde Z}(0,{\frak D}_{\chi})=
\pi \sqrt{-1}\eta(0,{\frak D}_{\chi})$.
It follows that

$$
{\widetilde Z}(0,{\frak D}_{\chi})=
{\widetilde Z}(0,{\frak D})^{{\rm dim}\,{\chi}}e^{-2\pi \sqrt{-1}CS(\chi)}
\mbox{,}
\eqno{(4.7)}
$$
and eventually the Chern--Simons functional takes the form

$$
CS(\chi)\equiv 
\frac{1}{2\pi \sqrt{-1}}{\rm log}\,\left[\frac{{\widetilde Z}(0,{\frak D})^
{{\rm dim}\,\chi}}{{\widetilde Z}(0,{\frak D}_{\chi})}\right]
\,\,\,\,{\rm mod}({\Bbb Z}/2)
\mbox{.}
\eqno{(4.8)}
$$
The classical factor becomes

$$
\exp\left[\sqrt{-1}kCS(\chi)\right]=\left[
\frac{{\widetilde Z}(0,{\frak D})^{{\rm dim}\,\chi}}
{{\widetilde Z}(0,{\frak D}_{\chi})}
\right]^{\hbar}
$$
$$
\times
\exp[2\pi \sqrt{-1} \hbar({\rm mod}({\Bbb Z}/2))]
\mbox{.}
\eqno{(4.9)}
$$

\subsection{The Ray-Singer norm}
For odd-dimensional manifold the Ray-Singer norm is a topological invariant: it
does not depend on the choice of metric on $X$ and $\xi$, used in the
construction. But for even-dimensional $X$ this is not the case
\cite{bism92}. For real hyperbolic manifolds of the form 
$\Gamma\backslash {\Bbb H}^3$ the dependence of the $L^2-$ analytic torsion 
(1.8) on zeta functions can be expressed in terms of Selberg functions
$Z_{\Gamma}(s;\chi)$. In the presence of non-vanishing Betti numbers 
$b_i\equiv b_i(X)= {\rm rank}_{\Bbb Z}H_i(X_\Gamma;{\Bbb Z})$) we have
\cite{byts1,byts4}

$$
\!\!\!\!\!\!\!\!\!\!\!\!\!\!\!\!\!\!\!\!\!\!\!\!
\!\!\!\!\!\!\!\!\!\!\!
[T_{an}^{(2)}{X}]^2=\frac{(b_1-b_0)![Z_{\Gamma}^{(b_0)}(2;\chi)]^2}
{[b_0!]^2Z_{\Gamma}^{(b_1-b_0)}(1;\chi)}
$$
$$
\times
\exp\left(-\frac{1}{3\pi}{\rm Vol}(\Gamma\backslash G)\right)
\mbox{.}
\eqno{(4.10)}
$$
There is a class of compact sufficiently large hyperbolic manifolds
which admit arbitrary large value of $b_1(X)$. Sufficiently large manifold 
contains a surface $\Sigma$ whereas $\pi_1(\Sigma)$ is finite and
$\pi_1(\Sigma) \subset \pi_1(X)$. In general, hyperbolic manifolds have not
been completely classified and therefore a systematic computation is not
yet possible. However it is not the case for sufficiently large manifolds
\cite{haken}, which give an essential contribution to the torsion (4.10).

Finally we note that formulae (4.8), (4.9), (1.2), (1.5) and (4.10) give the 
value of the asymptotics of the Chern--Simons
invariant in the one-loop expansion. The invariant involves the $L^2-$ 
analytic torsion on a hyperbolic 
3-manifold, which can be expressed by means of the Selberg zeta function 
$Z_{\Gamma}(s;\chi)$ and  
Shintani function ${\widetilde Z}(0,{\frak D}_{\chi})$, associated 
with the eta invariant of Atiyah--Patodi--Singer.

\end{document}